# Black Carbon scavenging in liquid Arctic clouds: the role of size and mixing state

Barbara Bertozzi[1,2]*, Robin L. Modini[1], Radovan Krejci[3,4], Gabriel Pereira Freitas[3,4], Rosaria E. Pileci[1,5], Paul Zieger[3,4], Martin Gysel-Beer[1]*

[1] PSI Center for Energy and Environmental Sciences, 5232 Villigen PSI, Switzerland
[2] Now at: Asterisk Labs, 86-90 Paul Street, London, England, EC2A 4NE
[3] Department of Environmental Science, Stockholm University, Stockholm, Sweden
[4] Bolin Centre for Climate Research, Stockholm University, Stockholm, Sweden
[5] Now at: STEP Srl, Milan, Italy
* Corresponding authors

## Abstract
Black carbon (BC) contributes to Arctic warming by absorbing sunlight and darkening snow. Its atmospheric lifetime critically determines its concentration and climate impact, yet the processes controlling its removal remain poorly constrained in the Arctic. From 18 months of single-particle measurements at the Zeppelin Observatory (Svalbard), we analysed 37 liquid cloud events (~200 hours) to investigate the link between BC properties and in-cloud scavenging, providing the most extensive in-cloud single particle BC dataset to date. While large BC cores ($D_{rBC}$>200 nm) were consistently scavenged, smaller cores were only partly removed. However, even thin soluble coatings significantly enhanced their scavenging, showing that mixing state modulates BC scavenging in the CCN-limited regime typical of Arctic low-level clouds. Seasonal variability in clear sky BC mixing state further suggests corresponding changes in scavenging efficiency. Our results demonstrate that besides size, the size-resolved BC mixing state is a key variable for BC scavenging in the Arctic and models should take it into consideration to accurately predict BC-cloud interaction.

## Introduction

Black carbon (BC) is a short-lived, light-absorbing aerosol component that plays a significant role in the Earth's radiative budget and climate. BC is a result of incomplete combustion and its properties vary with combustion source: urban emissions typically produce smaller, externally mixed and insoluble particles, while biomass burning tends to generate larger, internally mixed BC particles[1]. During atmospheric transport, BC undergoes aging processes such as coagulation, condensation, and cloud processing, that modify size, morphology, solubility, and optical properties of BC-containing particles. Accurately characterizing the size distribution and mixing state of BC throughout its atmospheric lifecycle is a necessary step toward understanding its interactions with clouds and removal processes[2]. However, a process-level understanding of how BC properties influence scavenging remains incomplete, limiting our ability to correctly represent BC removal in models[2,3]. This is particularly critical in the Arctic, where BC not only affects atmospheric radiation balance but also reduces surface albedo when deposited on snow[4,5].

Due to its high sensitivity to aerosol forcing, the Arctic has been the focus of extensive field and modelling studies in recent decades[6]. The seasonal cycle of Arctic aerosol is well documented: pollutant accumulation, including BC, during winter and early spring leads to the so-called Arctic Haze[7], while cleaner conditions dominate in summer, driven also by more efficient wet removal during long range transport[8]. Local and regional Arctic BC sources include shipping, gas flaring, residential heating, energy production and boreal wildfires[6]. Long-range transport is more efficient during the winter months and predominantly involves transport from Eurasia[6]. Recent studies have advanced beyond bulk BC concentration measurements to include particle size distributions, mixing state, and their corresponding vertical profiles. These single-particle datasets allow for a more accurate understanding of process-level phenomena, including the relationship between BC particle properties and their scavenging. The critical role of these parameters (i.e., size and mixing state) is evident when models try to simulate transport and processing of BC: tuning these parameters in one location causes mismatches in BC concentrations elsewhere[3].

Understanding BC scavenging mechanisms is crucial for proper assessment of BC lifetime and radiative effects, but it remains challenging due to limited observational datasets. Recently, a 4 years-long Arctic dataset has revealed a seasonal variability in BC bulk scavenging, with enhanced scavenging during summer and autumn[9]. Scavenging comprises two primary mechanisms: in-cloud scavenging via droplet-particles collision and coalescence, and nucleation scavenging, where particles serve as cloud condensation nuclei (CCN)[10]. Particle activation to CCN critically depends on cloud peak supersaturation ($SS_{peak}$) and particle properties, notably size and solubility that determine the particle critical supersaturation ($SS_{crit}$). Scavenging of small BC cores has been reported for high $SS_{peak}$ in previous in-cloud measurements[11–13], and the interplay between $SS_{peak}$ and $SS_{crit}$, and the consequent preferential scavenging of small BC cores when coated with soluble material, was shown in measurements at a high-altitude mountain site[13].

To better understand the properties of BC in the Arctic, several studies have investigated BC size distributions and mixing state using single-particle techniques. For instance, seasonality in BC size has been observed in the Canadian Arctic[14], with larger mass modal diameters during winter–spring (~225 nm) compared to summer (~170 nm). However, Arctic aircraft campaigns conducted between 2009 and 2017 found no clear seasonal pattern, with median core sizes in the range of $D_{rBC}$ = 202-210 nm[15]. Other Arctic studies report similar size ranges[16–19], typically larger than urban BC (which peaks around 140-150 nm)[12,20]. Comparisons of mixing state across studies remain challenging due to methodological differences and differences in BC size range investigated. Currently, the longest Arctic mixing-state dataset covers the winter period 2011-2012 and reported predominantly thickly coated BC (mean particle-to-core diameter ratio ~2.0)[18]. Other springtime Arctic studies have similarly observed particle-to-core diameter ratios ranging from ~1.5 to ~2.2[17,21].

In this study, we present size-resolved and mixing state-resolved BC measurements for both total and cloud residual aerosol from April 2019 to October 2020. Data partially overlap with the Ny-Ålesund Aerosol Cloud Experiment (NASCENT), which took place in Ny-Ålesund (Svalbard)[22]. Measurements were performed at the Zeppelin Observatory (474 m a.s.l.) at Ny-Ålesund Research station, Svalbard. The site is minimally affected by local sources, ideal for studying the properties of regional and long-range transported aerosols[23]. The observatory is often within clouds, making it a unique facility for studying aerosol-cloud interactions. Measurements with a ground-based counterflow virtual impactor (GCVI) inlet have been continuously performed at this site since 2015 to sample cloud residual particles that were e.g. analysed for their chemical composition[24], their single-particle composition and mixing state[25], or their biological properties[26]. A two-year dataset showed that activation diameters (i.e., the diameters corresponding to 50% of the aerosol population being scavenged into cloud droplets) on average ranged from ~58 to ~78 nm[27], indicating that cloud peak supersaturations are typically high. This result is confirmed by one year cloud modelling results showing $SS_{peak}$ values up to ~1% in fall and ~0.5% in summer[28]. These activation cut-off diameters are lower and corresponding cloud peak supersaturations are higher than typical values observed at a high-altitude site[29].

BC bulk scavenging at this site has been previously investigated for the period 2015-2019, revealing a distinct seasonal pattern: low wintertime scavenging and higher values in late spring and summer[9]. In addition, it was also shown in the same study that the scavenged fractions positively correlated to the cloud water content and that they decreased at lower temperatures, indicating the potential effects of in-cloud processes[9]. In this work, we expand on that by using a Single Particle Soot Photometer Extended Range (SP2-XR) to resolve size and mixing state of individual BC particles. By analysing both total (i.e., cloud residual and interstitial aerosol particles) and cloud residual BC, we investigate how BC properties relate to their scavenging behaviour in liquid clouds. Specifically, we address the following research questions: what is the monthly- and cloud-resolved variability in BC scavenging during the period covered by our dataset? Can meteorological conditions, cloud dynamics, and BC characteristics help explain the observed variability in BC scavenging efficiency in liquid Arctic clouds?

# Results

## Seasonal BC scavenging

In the period April 2019 – September 2020 we sampled total and cloud residual aerosol particles during 120 cloud events. Most of the sampled cloud events occurred during 2019 and lasted between 90 and 180 minutes, with the longest event lasting 25 hours (Figures S1a and S1b in the Supplementary Information, SI). We classify the sampled cloud events as either mixed-phase or liquid clouds based on their temperature, using -5 °C as the threshold. This value is motivated by findings from Tobo et al.[30], who showed that, in the Arctic, some aerosol particles can initiate ice nucleation at temperatures as high as –5 °C. Using this threshold, the clouds assigned to the liquid cloud group are unlikely to include mixed-phase clouds, whereas the clouds assigned to the mixed-phase group may include a few super-cooled liquid clouds and also pure ice clouds. Cold clouds occurred during the winter months, whereas the majority of sampled events are classified as liquid clouds (Figure S1c in the SI). The median value of the cloud-averaged visibility for the sampled clouds is ~175 m (Figure S1d). Most of the sampled clouds occurred under very clean conditions characterized by low total aerosol particle concentration (Figures S1e and S1f). For the total aerosol population (including interstitial particles and cloud residuals) the median cloud-averaged BC-free number concentration in the diameter range [100, 499) nm is 15.0 cm$^{-3}$ (min: 0.4 cm$^{-3}$, max: 115.3 cm$^{-3}$) and the median cloud-averaged rBC (refractory BC) mass concentration in the rBC mass equivalent diameter ($D_{rBC}$) range [68, 332) nm is 0.5 ng m$^{-3}$ (max: 18.9 ng m$^{-3}$). We refer to particles as BC-free when they produce a detectable scattering signal in the SP2-XR but do not exhibit a measurable incandescence signal, indicating the absence of a detectable amount of black carbon (more details on the instrument and the measured quantities are provided in the Method Section). For a complete description of the cloud-averaged properties of the sampled clouds, see Section 1 in the SI.

To simplify the interpretation of the results and to guarantee high data quality, we focus our analysis on a subset of 37 cloud events covering the period from April to September 2019 (i.e., from the end of the Arctic Haze to early autumn). This subset includes only cloud events that occurred under liquid cloud conditions to minimize cases possibly influenced by ice particles, which can affect GCVI inlet sampling efficiency and initiate the Wegener–Bergeron–Findeisen process, as already shown at this location[9,27] and elsewhere[11,31]. Additionally, events with extremely clean atmospheric conditions (i.e., BC-free particle number concentration lower than 10.0 cm$^{-3}$ or rBC mass concentration lower than 0.5 ng m$^{-3}$) are excluded, as limited counting statistics could compromise data quality. A more detailed description of the selection conditions is presented in the Method Section. The properties of the selected clouds directly reflect the filtering criteria resulting in higher median BC-free number concentration (45.2 cm$^{-3}$) and rBC mass concentration (1.4 ng m$^{-3}$) in the total aerosol population. Figure S1 in the SI provides a complete comparison of the cloud-averaged properties of the sampled and selected clouds.

By dividing the selected clouds by month, we look for monthly variability in the properties of BC particles sampled from the total aerosol population during the selected events (Figure 1a) and investigate if and how seasonality influences the scavenging of BC particles in liquid arctic clouds (Figure 1b). The monthly distributions of cloud-averaged total-inlet rBC bulk mass concentration (i.e., for $D_{rBC} \in$ [68, 332) nm) follow the expected trend with higher values at the end of the spring and cleaner conditions during summer (white boxes in Figure 1a). The division between spring and summer is indeed defined by the change in aerosol properties between the Arctic Haze (spring) and the cleaner summer conditions[32]. For the events of April 2019, the median cloud-averaged rBC mass concentration is 2.4 ng m$^{-3}$, which is higher than the overall median value reported above. During the selected summer cloud events (occurred in May-Aug), instead, the rBC mass concentration is lower, with medians in the range from 1.1 to 1.4 ng m$^{-3}$. This seasonal pattern is consistent with the multi-year in-cloud seasonality of equivalent black carbon mass concentration reported at the same location[9].

The variability observed in the size distribution boxplots of Figure 1a might reflect both changes in concentration and differences in the shape of the rBC mass size distributions within individual months. The normalized cloud-averaged mass size distributions underlying these boxplots are shown in Figure S2 in the SI. The mass size distributions during the selected cloud events do not exhibit a clear seasonal trend, with 90% of the modal diameters falling within the range [130, 232) nm. July 2019 shows the widest spread in modal diameters, suggesting greater variability in rBC mass size distributions during that month. Since the peak of the BC particle number size distribution lies below the lower $D_{rBC}$ detection limit of the SP2-XR (Figure S3), modal diameters derived from lognormal fits to number distributions are less reliable and thus not directly compared here.

The obtained bulk rBC mass and number scavenged fractions ($F_{rBC}$) are relatively high with median values of all the selected cloud of $F_{rBC,mass} = 0.85$ (inter quantile range IQR: 0.68-0.90) and $F_{rBC,numb} = 0.81$ (IQR: 0.61-0.90). The slightly lower bulk scavenged fraction based on number concentration compared to mass concentration reflects differences in the respective size distributions. The number size distribution is dominated by smaller particles, while the mass distribution is weighted toward larger diameters. Although our data only cover the end of the spring and the summer 2019, we obtain monthly mass scavenged fractions in agreement with the seasonal trend based on four years 2015-2019 [9], showing almost full scavenging in early summer (May-June) and less scavenging in spring and late summer (April and July-August). The lowest cloud-averaged mass scavenged fraction of 0.42, for the liquid clouds discussed here, was measured for a cloud event that occurred in July 2019. In addition to month-to-month variability, there is also intra-month differences in bulk scavenged fractions across cloud events (grey and white boxes in Figure 1b). April and May 2019, for example, have a much lower variability between cloud events compared to the summer months of 2019.

In addition to the bulk scavenged fractions, the SP2-XR provides size-dependent scavenging data that can help to explain observed bulk scavenged fractions and their variability. Months with the lowest variability in the bulk scavenged fraction (April and May) are also the months for which the scavenged fractions are essentially independent of size (as shown in Figure 1b with purple colours). July and August, on the contrary, have low median bulk scavenging fraction and high variability resulting from a large variability in the scavenging of small BC particles. The lower scavenged fractions and higher variability of the cloud events from June reflect a transition from the full scavenging at all sizes to a size-dependent scavenging behaviour. It is important to note that the size-resolved scavenged fractions are, by definition, identical for BC mass and BC number. However, since the smallest BC particles detected by the SP2-XR are larger than the majority of the BC particles present in the atmosphere (i.e., the peak of the number size distribution is below the detection limit of the SP2-XR), the scavenged fraction calculated for the smallest size bin can be considered an upper limit for the overall bulk number-based scavenged fraction.

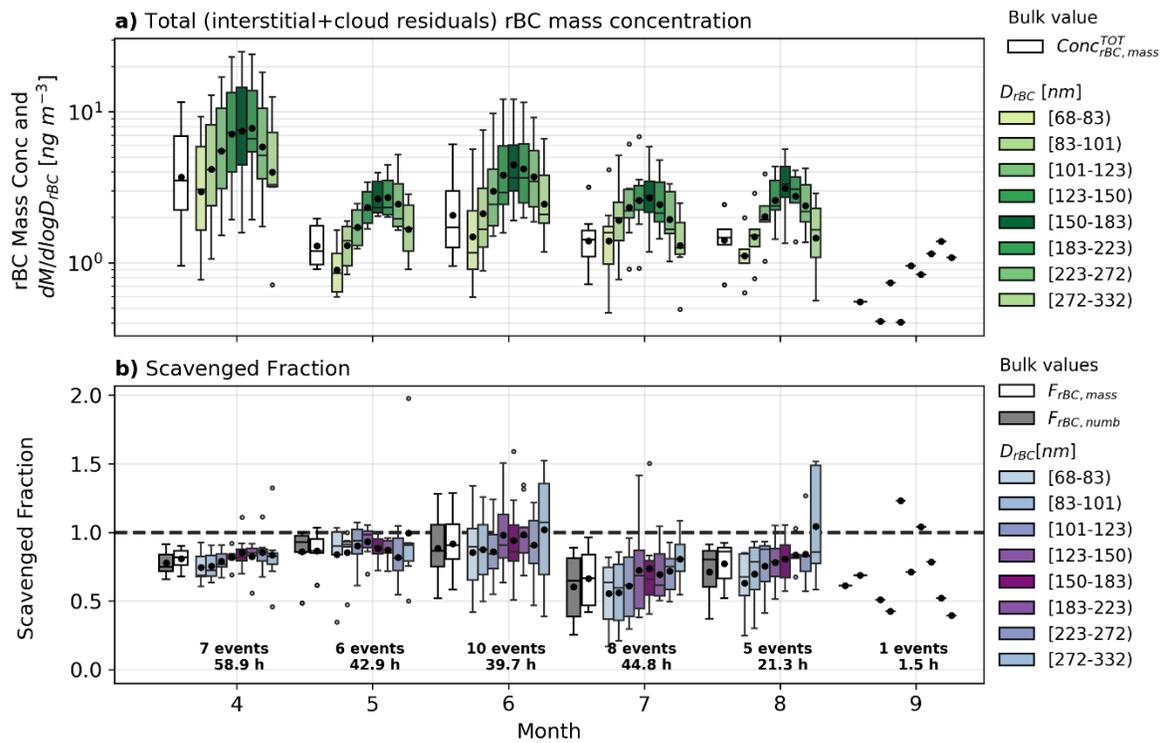

*Figure 1 **Distribution of cloud-averaged BC data for the selected cloud events grouped by month.** (a) Bulk mass concentration ($Conc_{rBC,mass}^{TOT}$, white) and mass size distribution (green colours) sampled from the total aerosol population (i.e., interstitial aerosol and cloud residual), (b) bulk mass ($F_{rBC,mass}$) and number ($F_{rBC,numb}$) scavenged fractions (white and grey, respectively) and size dependent scavenging curves (purple colours). Bulk BC values are obtained via integration over the range $D_{rBC} \in [68, 332)$ nm. In this and in the following plots, boxes show the interquartile ranges, and the whiskers represent the lower and upper quartiles. Mean and median are represented with a black circle and a horizontal line, respectively. Outliers represent points below and above the lower and upper quartiles. Number of cloud events and total number of hours selected for each month are also reported.*

### Different BC scavenging regimes

The measured size resolved BC scavenging indicates that small BC particles are often not fully scavenged into cloud droplets in low level liquid Arctic clouds, despite the expected high cloud peak supersaturations[27] and the aging of the black carbon particles transported to such a remote location. We have thus investigated the factors and properties contributing to differences in BC scavenging by grouping the size-resolved scavenging curves $F_{rBC}(D_{rBC})$ according to their shape and values.

We classify the BC scavenging curves as "fully scavenged" (type A) when more than 90% of the BC particles with a BC core size in the range $D_{\mathrm{rBC}} \in [68,83)$ nm are scavenged. The remaining cloud events, which exhibit size dependent activation, are further divided in two additional classes by means of a clustering algorithm based only on the size-resolved scavenging curves. More details on the classification are given in Section 3 in the SI. The statistics of the cloud-averaged scavenging curves $F_{rBC}(D_{rBC})$ for the three groups, as determined by the classification scheme, are presented in Figure 2. Group A (10 cloud events) includes cases where BC particles are completely scavenged across the probed size range, indicating that scavenging is independent of BC properties. The remaining events, which exhibit a size-dependent scavenging behaviour, are further categorized by the scheme into Group B (17 cloud events), showing a weak size dependence, and Group C (10 cloud events), characterized by a strong size dependence. The $F_{rBC}(D_{rBC})$ curves are clearly separated between each class at all sizes. However, differences are more prominent for small BC cores. The variability of $F_{rBC}$ in the range $D_{rBC} \in [272, 332)$ nm is amplified by noise due to poor counting statistics in this size range (as also visible from the BC number size distributions in Figure S3). All the cloud-averaged scavenging curves are reported in Figure S4 in the SI.

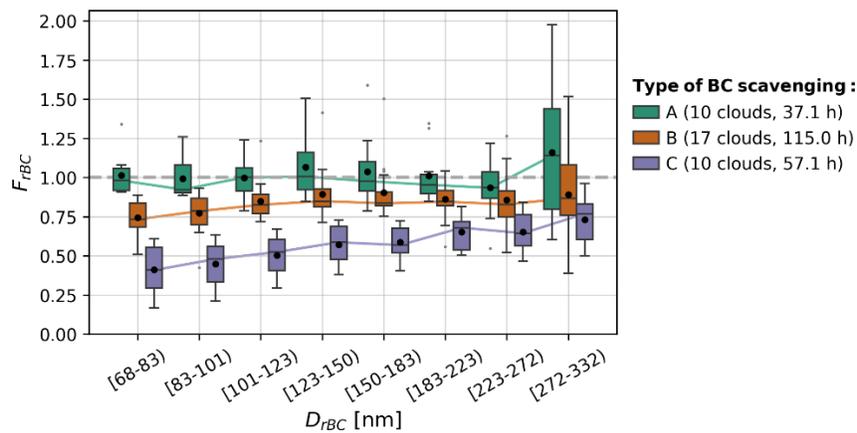

Figure 2 *Size-resolved BC scavenging, $F_{rBC}(D_{rBC})$, for the selected cloud events grouped based on their BC scavenging type (colors)*.

We interpret a steeper gradient in the size-resolved $F_{rBC}$ and reduced values at small $D_{rBC}$ as indicative of stronger selectivity in the nucleation scavenging process with respect to BC particle properties. This classification can therefore serve as a basis for exploring the factors that control the selective nucleation scavenging of BC particles. As already evident from Figure 1b, there is a link between scavenging type and month. Type A events occurred only in April, May, and June. Type B events, with weak size dependence, are more uniformly distributed throughout the observation period, while events with scavenging of type C occurred mostly in June, July, and August. Figure S1 in the SI provides an overview of the distribution of cloud-mean properties such as duration, temperature, and visibility for all the sampled clouds and for the selected cloud events across the different scavenging types. In the following, we focus on some of the mentioned properties to investigate possible connections with the three types of BC scavenging curves. We investigate the role of (i) cloud dynamics (i.e., cooling rate), (ii) properties of the BC-free aerosol population, and (iii) properties of the BC-containing particles.

Having already excluded cold cloud events, the selected cases span a relatively narrow temperature range, from -4.7 °C to 5.6 °C. Despite this limited range, the seasonal pattern described above is reflected in a temperature trend across the three BC scavenging types (Figure 3a). Events of type A have median temperature of -3.0 °C, -0.3 °C for type B, and 2.6 °C for type C. For all selected events, the cloud-mean temperatures are sufficiently high to exclude interference of mixed phase clouds to the group of liquid clouds assessed here. All selected clouds have a cloud-averaged visibility lower than 250 m ensuring that measurements were carried out deep inside the clouds minimizing the effect of partial activation near the cloud edge or high variability due to changes in cloud density. Figure 3b doesn't show any clear correlation between cloud visibility and BC scavenging type, which is also not expected for a droplet nucleation process, except for samples directly at cloud base with very low liquid water content and very high visibility.

Since cloud peak supersaturation was not directly measured during our campaign, we use horizontal wind speed and concentration of BC-free particles as proxies to assess the cloud formation regimes. At the Zeppelin Observatory, updraft, is largely driven by orographic forcing making horizontal wind speed a useful proxy for the cooling rate. At the same time, the number of available cloud condensation nuclei (CCN) determines how quickly the excess water vapor is depleted during initial stages of cloud droplet formation at cloud base. When CCN concentrations are low compared to updraft, the excess water vapor is not fully depleted by the small number of formed and growing cloud droplets. Instead, it remains in the gas phase higher up above cloud base, allowing the air parcel to reach higher peak supersaturation, before the increasing condensation sink offered by the grown droplets eventually depletes the excess water vapor (the so-called CCN-limited regime[33]). Here, we consider the ratio of updraft strength (approximated by horizontal wind speed) to CCN availability (approximated by the concentration of BC-free particles in the [100, 499) nm diameter range) as a qualitative indicator of the cloud formation regime. Figure 3c shows that type A cloud events are associated with higher wind speeds, while Figure 3e shows that they also tend to have lower BC-free concentrations. These conditions yield a relatively high updraft-to-CCN ratio, consistent with formation in a strongly CCN-limited regime. In contrast, type B and C events show slightly lower wind speeds and higher BC-free concentrations, pointing to a less CCN-limited regime with slightly lower peak supersaturations. This interpretation that cloud formation typically occurs in the strongly or partly CCN-limited regime is consistent with previous observations and modelling studies at the Zeppelin Observatory[27,28].

Additional evidence of the influence of cloud peak supersaturation on BC scavenging comes from the scavenged fraction of BC-free particles in the [100, 150) nm diameter range (Figure 3d). Among the selected cloud events, the lowest measured cloud averaged $F_{BC-free}$ in the mentioned diameter range is 0.72, meaning that for all the selected clouds the 50% activation cut-off diameter for BC-free particles is lower than 100 nm, which indicates high peak supersaturations. Furthermore, median $F_{BC-free}$ values for the same diameter range are lower for type B (0.89) and type C (0.93) events than for type A (1.00), supporting the hypothesis that the cloud events with lower BC scavenged fraction experienced somewhat lower peak supersaturations. Stronger size dependence of BC free particle scavenged fraction for type B and C compared to type A clouds further supports the interpretation of systematic supersaturation differences (Figure S5 in the SI). A similar link between cloud supersaturation and size-resolved BC scavenging has been previously observed in other locations[13]. In addition to dynamic conditions and particle concentrations, BC scavenging into cloud droplets also depends on their intrinsic properties like mixing state. In the following section, we examine the role of BC mixing state for the three cloud groups.

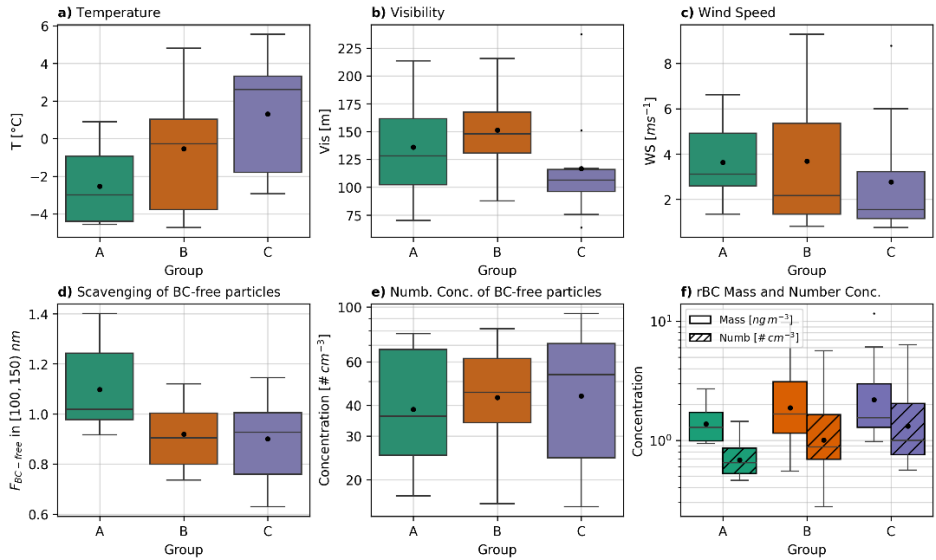

*Figure 3 **Distributions of various meteorological conditions and aerosol properties during the different cloud events grouped by BC scavenging types.** Temperature, visibility, and wind speed data were measured on the rooftop of the Observatory. The scavenging of BC-free particles (panel d) refers to the diameter range [100, 150) nm. The number concentration of BC-free particles (panel e) is calculated across the diameter range [100, 499) nm. The number and mass concentrations of BC particles (panel f) refer to the diameter range $D_{rBC} \in$ [68, 332) nm.*

## Mixing state dependence of BC scavenging

Mixing state of BC with other particulate matter is an important parameter because it affects size and solubility of BC-containing particles. Measuring the fraction of internally mixed particles and the volume fraction of coating material is not trivial but nevertheless important to correctly represent black carbon radiative effect and lifetime. In this study we use the time delay method[34] applied to SP2-XR data to qualitatively classify BC particles with $D_{rBC}$ larger than ~100 nm according to their mixing state. This method distinguishes, on a single particle level, between BC particles with low ("uncoated to moderately coated" BC) and with medium to high BC volume fraction ("thickly coated" BC). For the smallest BC cores, it is possible to distinguish a further subgroup of "bare or almost bare" BC particles among the uncoated to moderately coated group. A detailed description of how the time delay method is implemented is presented in Section 5 in the SI.

To evaluate the role of BC mixing state in defining the BC scavenging behaviour, we compare the mixing state of BC particles in the total aerosol population across the three scavenging types. Since the event classification is based solely on the size-resolved scavenging of BC, the mixing state provides an independent variable that can help the interpretation. Figure 4 shows the cloud-averaged size-resolved fraction of thickly coated BC particles in the total and cloud residual aerosol populations, grouped by scavenging type. The differences in mixing state are especially pronounced in the total population for $D_{rBC}$ smaller than ≈150 nm. Events classified in group A, characterized by complete scavenging across all sizes, correspond to periods with a higher fraction of thickly coated BC particles (median fraction of 0.71 for $D_{rBC} \in$ [101, 123) nm). In contrast, type C events, showing the strongest size dependence, exhibit the lowest fraction of thick coatings (median fraction of 0.43 for $D_{rBC} \in$ [101, 123) nm). These differences suggest a combined influence of cloud peak supersaturation (as discussed above) and BC mixing state on the BC scavenging behaviour. Particles with equal rBC mass but higher fraction of coating material are more frequently scavenged in liquid Arctic clouds. In comparison, the fraction of thickly coated BC particles in the cloud residuals is consistently higher than for the total aerosol and it shows less variation across cloud groups, with median values of 0.72, 0.65, and 0.57 for $D_{rBC} \in [101, 123)$ nm for groups A, B, and C, respectively (Figure 4). This highlights that BC particles with small core diameters are preferentially scavenged when they are internally mixed, which supports the interpretation that BC mixing state plays a significant role in nucleation scavenging.

The concentrations of rBC mass and BC particle number increase from group A to group C clouds. However, we do not interpret this as a causal link between BC concentration and BC scavenged fraction. Instead, this may just be a cross-correlation. Specifically, some causal link between the ratio of BC-containing to BC-free particles, which increases from group A to group C clouds based on the trends in Figures 3e and 3f, and the BC coating fraction is expected through the availability of non-BC matter relative to BC. Also, it has been previously pointed out that decreasing BC scavenged fraction with increasing BC concentrations likely was the result of cross-correlations rather than direct causality[11].

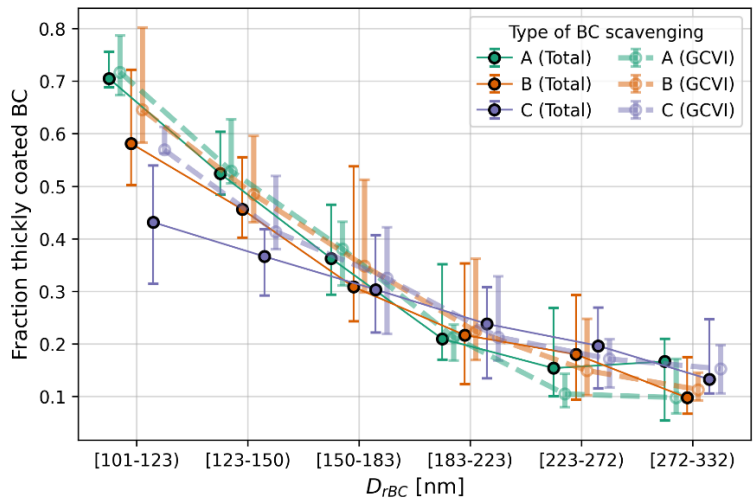

*Figure 4 Size-resolved fraction of thickly coated BC particles in the total and cloud residual aerosol populations grouped by the three BC scavenging types. The fraction of thickly coated BC particles present in the total aerosol population is shown with solid dots and lines; the fraction of thickly coated BC particles present in the cloud residual population is shown with semi-transparent dots and dashed lines. Colours refer to the different BC-scavenging groups. The clear separation between the thickly coated fraction of the three BC scavenging groups for the total aerosol population (for $D_{rBC} < 150$ nm) demonstrate the link between the BC mixing state and its cloud scavenging. The consistent higher fraction of thickly coated BC particles in the cloud residual population supports the hypothesis that scavenging of small BC cores is selective to its mixing state.*

To further assess the impact of BC mixing state on BC scavenging, we evaluate the size-resolved BC scavenged fractions observed for each cloud group separately for the subsets of thickly coated, uncoated to moderately coated, and externally mixed BC particles. This approach isolates the role of BC mixing state by comparing the scavenging of particles with different mixing states under similar environmental conditions, thereby disentangling its effects from the role of cloud peak supersaturation. Figure 5 shows BC scavenging curves decomposed in thickly coated (black boxes) and uncoated to moderately coated (coloured boxes) BC particles for the three cloud groups (panels a, b and c). The subgroup of bare or almost bare BC particles is also shown (white boxes). For all three cloud groups (A, B, and C), the thickly coated particles represent the ones with highest and almost complete scavenging. For type B and C cloud events, the uncoated to moderately coated BC particles (coloured boxes) exhibit significantly lower scavenged fraction especially at the smaller size bins, thus further demonstrating mixing state dependence of BC scavenging. This is further corroborated by the externally mixed BC subgroup (white), which has very low scavenged fraction for all clouds.

It is important to recall that the time delay method gives a simplified, binary classification of BC mixing state for a given core size, i.e. coating below/above a threshold. However, we can assume that coating thicknesses form a continuous distribution, meaning that within the group "uncoated to moderately coated" particles may still include a range of coating amounts. As a result, when the fraction of thickly coated particles is high (as in type A clouds), the remaining uncoated to moderately coated ones are likely to have, on average, thicker coatings than in clouds with fewer thickly coated particles (like in types B and C). Therefore, the group classified as "uncoated to moderately coated" differs in its average coating level between cloud groups. This is important because we observe that BC particles in this group are scavenged less efficiently from type A to type C events. But this trend may not be due only to changes in cloud peak supersaturation, it could also reflect actual differences in BC mixing state between the cloud groups.

For all three BC scavenging types, the scavenging of large, thickly coated BC particles is consistently below unity and shows greater variability compared to the scavenging of smaller thickly coated particles or BC particles of the same size that are uncoated to moderately coated. This increased variability is likely due to differences in the absolute concentration of uncoated to moderately coated and thickly coated particles in this size range. Indeed, for larger core diameters, most particles are classified as uncoated to moderately coated, making the scavenging fraction for thickly coated ones more susceptible to statistical noise.

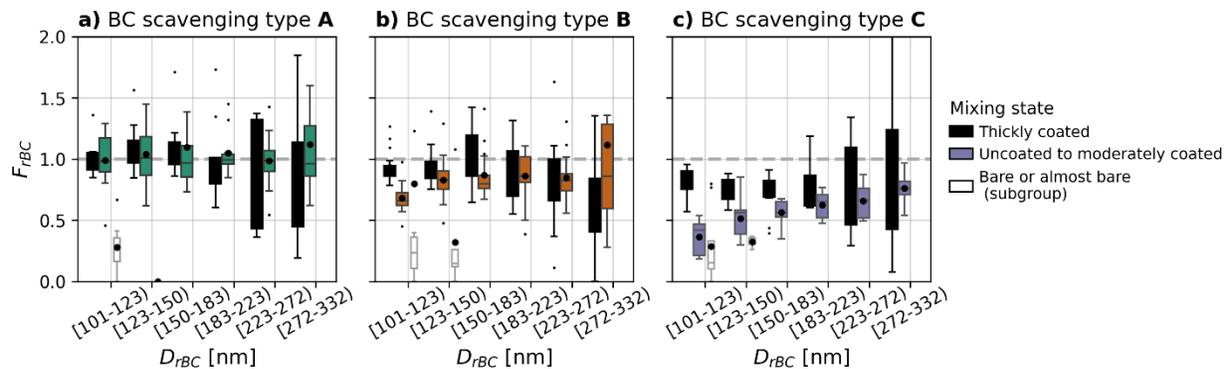

*Figure 5 **Size and mixing state resolved BC scavenging.** The identified three groups of BC scavenging curves (panels a, b, and c) are decomposed according to the BC mixing state (colour of the boxes). For all BC scavenging groups, the thickly coated BC particles (black boxes) are almost fully scavenged at all sizes. For events with BC scavenging of group B and C, the uncoated to moderately coated BC particles (coloured boxes) exhibit significantly lower scavenged fraction especially at the smaller size bins, thus further demonstrating mixing state dependence of BC scavenging.*

# Conclusions

## Discussion

The size and mixing state resolved measurements we performed in the high Arctic during liquid cloud events (April – September 2019) represent the longest single-particle in-cloud BC dataset to date. These measurements show that small BC particles are often only partially scavenged into cloud droplets. In contrast, larger BC cores are scavenged more efficiently, regardless of their mixing state. This highlights the crucial role of even a small amount of soluble coating in enabling small BC particles to activate as CCN, despite the high cloud supersaturations typical of Arctic clouds forming in a strongly CCN-limited regime.

While previous studies have explored how BC core size, mixing state, and cloud supersaturation affect scavenging efficiency, few have done so in the Arctic. Measurements at different geographical locations show that cloud supersaturation largely determines the BC critical activation diameter, while BC mixing state further modulates activation efficiency. Results most comparable to ours were obtained at the Jungfraujoch observatory, where increasing cloud supersaturation (0.2% - 0.5%) led to progressively higher BC scavenging and, for small BC cores, preferential activation of the thickly coated fraction. In contrast, lower supersaturation conditions, such as urban fog[12] ($SS_{peak}$ approx. 0.04%) or marine stratocumulus[35] (estimated $SS_{peak}$=0.1%), were associated with significantly larger critical BC core diameters (~200 nm) and very low scavenged fractions (1-10%), respectively.

In the Arctic, previous studies have focus on BC bulk scavenging[9] or on combining measurements of cloud residuals with total aerosol observations taken below and above cloud layers[36]. In both cases, interpretation is complicated either by a lack of mechanistic details or by the influence of additional processes like below cloud scavenging and entrainment. Our study addresses these limitations with semi- collocated, single-particle measurements of both total and cloud-residual BC in Arctic clouds, where aerosol sources, atmospheric dynamics, and aging pathways differ substantially from mid-latitude settings.

Although our setup does not allow a direct separation between nucleation and in-cloud impaction scavenging, the strong dependence of scavenging on both BC size and mixing state points to nucleation as the dominant mechanism for BC scavenging in liquid Arctic clouds. Nucleation scavenging depends on the interplay between cloud supersaturation and the particle's critical supersaturation, which is itself determined by particle size and composition.

## Implications

Our in-cloud measurements show that BC core size and mixing state significantly influence its scavenging efficiency in Arctic liquid clouds. However, the selected events cover a limited period and specific meteorological conditions. Previous studies have shown that BC sources and transport pathways at the Zeppelin Observatory vary with season and differ between cloudy and clear-sky periods[9]. To evaluate whether our in-cloud findings are representative, we compare them with the 18 months of clear sky SP2-XR measurements.

While part of this dataset was previously analysed in Pileci's PhD thesis[37], we extend the analysis here by including a longer time period and presenting the seasonality in BC mixing state. This property affects not only particle solubility, and thus its cloud scavenging, but also its optical properties. Past Arctic studies of BC mixing state have typically been limited in duration, therefore not addressing seasonal variability, and generally reported high particle-to-shell diameter ratios (1.5–2.2)[14,17–19,21].

Seasonal patterns appear both in BC size and mixing state. Monthly median mass size distributions (Figure 6a) show slightly larger BC cores in winter than in summer (171 nm in August 2019 and 224 nm in January 2020), consistent with earlier Arctic observations[14]. The mixing state also varies by season (Figures 6b and 6c), particularly for larger BC particles, which show a higher fraction of thickly coated particles in summer. Variability is highest in summer, with the interquartile range of thickly coated BC with $D_{rBC} \in [101, 123)$ nm spanning nearly the full range observed during cloud events. This increased variability may contribute to the larger spread observed in BC scavenging behaviour during summer (Figure 1b). The observed seasonal trend in BC mixing state likely reflects differences in source regions, transport pathways, and atmospheric processing of BC. However, identifying the dominant drivers will require further analysis, such as combining SP2-XR data with back-trajectory modelling[38].

Overall, the clear-sky results support the variability observed during the selected clouds and highlight the need to account for seasonal variability in BC mixing state, which influences not only scavenging efficiency but also radiative properties.

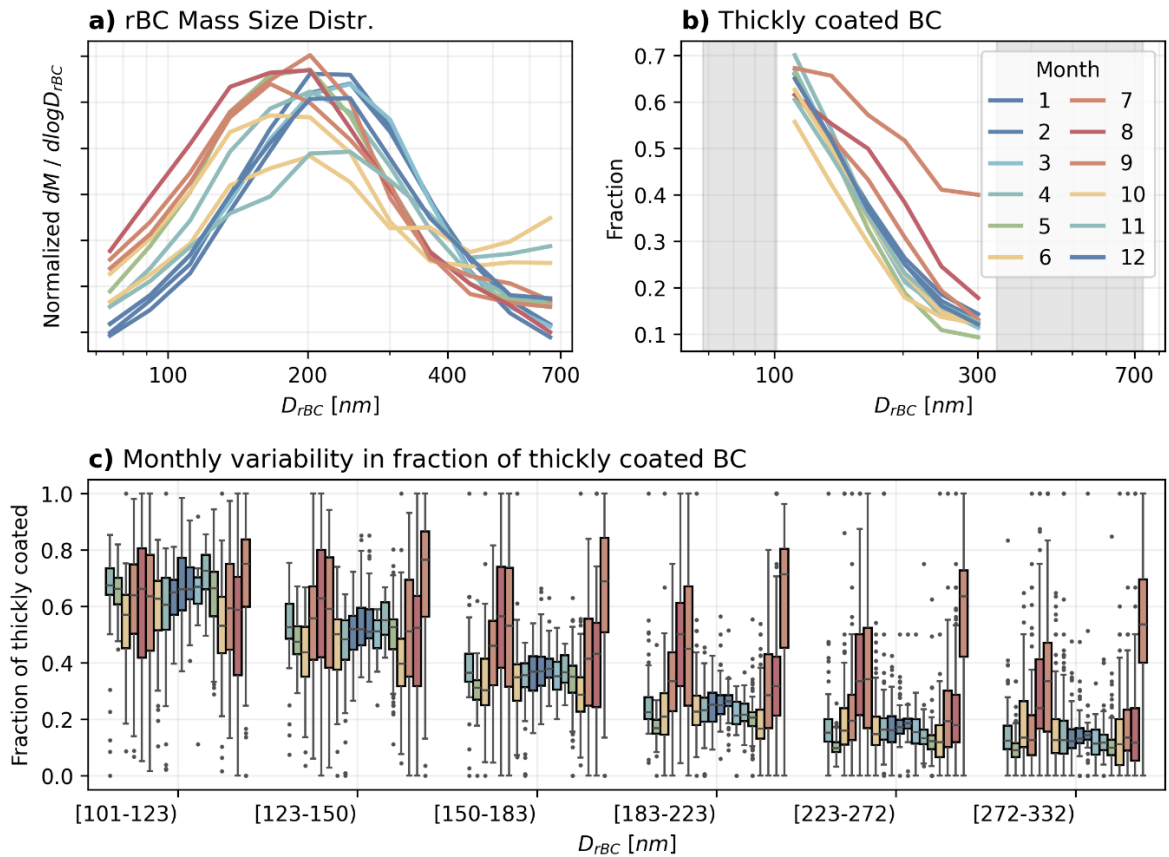

*Figure 6 **Seasonality of clear-sky BC measurements (5-hour time resolution) from April 2019 to September 2020**. Panel a: Monthly median rBC mass size distribution normalized by the corresponding mass concentrations calculated for $D_{rBC} \in [68, 734)$ nm to emphasize relative changes in shape. Panel b: Monthly median fractions of thickly coated BC particles. Panel c: Statistics of the size-resolved fraction of thickly coated BC particles. A seasonal trend is evident in the BC mass size distribution and mixing state. This variability may reflect differences in BC sources, transport pathways, and atmospheric processing.*

## Outlook

Our results show that overall BC scavenging in liquid Arctic clouds is high primarily due to the characteristic high cloud peak supersaturations. These conditions lead to the scavenging of most large BC cores, regardless of their mixing state. In contrast, the scavenging of smaller BC particles is more variable and selective, depending on both cloud supersaturation and particle mixing state. Thickly coated BC particles are more readily activated due to their increased total diameter and enhanced solubility from the coating material.

The observed dependence of small BC particle scavenging on mixing state and supersaturation, together with the seasonal patterns identified in the clear-sky dataset, suggests a corresponding seasonal variability in BC scavenging behaviour. This highlights the complex interplay between cloud dynamics and BC properties in determining the fate of BC in Arctic clouds. Importantly, our findings challenge the assumption that high supersaturation and aged aerosol always lead to near-complete BC scavenging.

These results have important implications for climate modelling. Current regional and global models often assume a fixed or oversimplified BC scavenging efficiency, which may not adequately capture the observed variability. To improve the representation of BC transport, aging, lifetime, and climate effects, models should incorporate schemes that account for variability in scavenging efficiency as a function of both cloud microphysics and size-resolved BC particle properties[39,40]. However, the development of such schemes must be guided by comprehensive experimental data that capture the diversity of atmospheric conditions, BC characteristics, and cloud regimes across seasons and regions.

## Methods

### Black Carbon Measurements

Black carbon measurements presented in this manuscript were obtained with the Single Particle Soot Photometer Extended Range (SP2-XR) instrument. The SP2-XR[41], a modified version of the SP2[34,42], measures the optical size of all sampled particles and the mass of particles that incandesce at the instrument's laser wavelength. The instrument is equipped with a 1064 nm wavelength laser to illuminate the sample flow, and two detectors to collect the light elastically scattered by each particle and the laser induced incandescence (LII) signal produced by the absorbing particles. Calibration of the two detectors allow to establish a relation between scattering and optical diameter when using a non-absorbing calibration material, and between incandescence signal intensity and mass of the refractory black carbon (rBC) when using a suitable bare BC as calibration material.

The SP2-XR ran continuously and mostly unattended at the Zeppelin Observatory from April 2019 till September 2020. The SP2-XR was calibrated on site in June 2019 and in November 2019 by sampling electrical mobility-selected ammonium sulphate particles in the mobility diameter range 80-500 nm for the scattering detector. Mobility-selected fullerene soot particles in the mobility diameter range 50-540 nm were used for calibration of the incandescence detector, thereby following recommendations and mobility density data available in the literature[43,44]. A material density of 1.8 g cm$^{-3}$ is assumed for the conversion from rBC mass in a single particle to the corresponding $D_{rBC}$. During the calibration, a condensation particle counter sampled in parallel to the SP2-XR to validate the instrument counting efficiency. As a result of the counting efficiency test, we only consider BC-free containing particles in the diameter range between 100 nm and 499 nm and BC-containing particles with $D_{rBC} \in [68, 734)$ nm. Due to the low concentrations measured at large $D_{rBC}$, we further restrict the in-cloud data to particles with $D_{rBC} < 332\ nm$ to reduce statistical noise in the scavenging fraction calculations. The rBC mass concentrations measured during the NASCENT campaign were systematically lower than collocated equivalent BC (eBC) mass concentration measurements[22], but within the range of uncertainty in BC mass concentration measurements[14,45].

The quality and continuity of the collected data, combined with the instrument's robustness, demonstrate its suitability for long-term, unattended operation even in demanding and remote environments. However, the length of these single-particle datasets poses significant challenges for data processing due to the large data volume and the resulting long computation times required for data aggregation. To address this, we developed a Python code using the Dask library[46], enabling parallel and distributed computation on the PSI high-performance computing system and substantially reducing processing time. Data were initially gridded with 1-minute time resolution and in eight and 19 lognormally spaced size bins for BC and BC-free particle size distributions. The bulk number and mass concentrations of BC particles as well as the bulk number concentration of BC-free particles have been calculated across the same diameter range used for the size distributions ([68, 332) nm for the BC in-cloud data and [68, 734) nm for the BC clear sky data; [100, 499) nm diameter range for the BC-free number concentration).

In addition to the BC size distribution, the SP2-XR measurements can provide valuable qualitative information on the mixing state of BC-containing particles with $D_{rBC} > 100$ nm. In particular, the time elapsed between the peak of the elastic scattering and the peak of the incandescence signal provides a qualitative information on the particle mixing state[34]. Section 5 of the SI presents a detailed description of the time delay method with examples from the current dataset (Figure S6).

## Experimental Set-Up

The measurements presented in this manuscript were performed at the Zeppelin Observatory (Svalbard) as part of the NASCENT campaign, whose overall experimental setup and goals are presented elsewhere[22]. The inlet set-up used in this study has been in operation for several years at the Zeppelin Observatory and has been validated in numerous previous studies, demonstrating its suitability for aerosol-cloud interaction studies[9,26,27]. A detailed description and characterization of the total and cloud residual inlets of the Zeppelin Observatory is given in Karlsson et al.[27], therefore only a brief introduction is provided below. Among the default instrumentation present at the Observatory we mention the temperature[47], visibility, and wind speed[47] sensors located on the rooftop of the Observatory and used in this manuscript to identify cloud events and characterize their meteorological conditions.

During clear sky and cloudy periods, a total aerosol inlet sampled the whole air aerosol population. The inlet line is heated to approx. 5-10°C and any dyer is used because the temperature difference with the external air dries the sampled flow when transported in the laboratory[27]. Hydrometeors are thus evaporated in the inlet and both interstitial and cloud residual particles are sampled by the instruments. Cloud residual particles are sampled with a ground-based counterflow virtual impactor inlet (GCVI, Brechtel Manufacturing Inc., USA, Model 1205). The GCVI inlet sampled hydrometeors larger than approx. ~6-7 μm, thus overall droplet sampling efficiency (SE) depends on the position of the cloud droplet size distribution relative to this cut-off[27]. The GCVI inlet operates as virtual impactor by opposing the sample flow with a counterflow of dry air. In this way only particles with a high inertia can win the counterflow and being sampled by the instruments. The working principle of the GCVI leads to an enrichment of initially sampled droplets in the finally extracted sample flow. The magnitude of this enrichment factor (EF) is calculated according to ambient conditions, instrument geometry and instrument operation. For the system operated at the Zeppelin Observatory and for the selected cloud events presented in this manuscript the cloud averaged EF was very stable ranging between 10.1 and 10.3. As in previous studies, we estimate the GCVI sampling efficiency (SE) by assuming complete scavenging of accumulation mode particles. Specifically, we match the number concentration of BC-free particles in the diameter range [254, 499) nm measured from the total and GCVI inlets (Figure S7). By using SP2-XR data of BC-free particles to derive SE at high time resolution (~20 minutes, Figure S8), we account for short-term variations in droplet size distribution and minimize the influence of potential differences in particle losses between the two inlets. The obtained SE values are in good agreement with values obtained for the same location in previous studies (Figure S9). A detailed description of the SE estimation is provided in Section 6 in the SI.

The SP2-XR was installed at the Zeppelin Observatory with an automated three-way valve connected to the total and GCVI inlets. When the GCVI was not in operation (e.g., during clear sky periods), the SP2-XR sampled continuously from the total aerosol inlet. When the GCVI was in operation, instead, the three-way valve automatically started switching between the two inlets every 10 minutes. The GCVI sampling efficiency calculated from the BC-free SP2-XR measurements have been calculated with a 20-minute time resolution and then applied to the 1-minute data. The in-cloud data presented in the manuscript are cloud-averaged values from the 1-minute corrected data. Cloud-averaged scavenging fractions have been calculated as the ratio of the corrected cloud-averaged residual concentration to the cloud-averaged total concentration. The clear sky data have been averaged to a 5-hour time resolution before computing the statistics shown in Figure 6 to maintain a comparable averaging time with the in-cloud data (the mean cloud duration of the selected cloud events is 5.7 hours).

## Cloud identification and selection

In this work, a cloud event is identified when the visibility measured on the rooftop of the Observatory is below 1 km for at least 90 minutes, and both GCVI and SP2-XR are in operation. During the campaign period, we collected data during 120 cloud events for a total of ~600 hours. Only a subset of clouds was then selected for further analysis to simplify the interpretation of the results (by excluding mixed-phase clouds), exclude extremely clean conditions (data interpretation hampered by poor counting statistics, see Figure S10), exclude cloud events for which the estimated GCVI sampling efficiency (SE) was below 20% (to exclude events with a poor representation of the cloud droplet distribution) and exclude events for which the relative difference in the GCVI-SE estimated in the diameter range [153, 254) nm and [254, 499) nm differ more than 0.5 (see Figure S11). In Section 7 of the SI, we present a complete description of the filtering criteria and show the corresponding effect on the distribution of cloud properties. In Figure S1 in the SI the distribution of the cloud-average properties for the sampled and selected clouds is shown. In Section 8 in the SI, a table summarizes some of the mean or median parameters relative to the selected cloud events analysed in the manuscript.

**Acknowledgements**

We acknowledge Günther Wehrle for his support in operating the SP2-XR instruments during the field measurements. The authors would like to thank the Norwegian Polar Institute (NPI) for their long-term support at the Zeppelin Observatory. We thankfully acknowledge the research engineers Tabea Hennig, Zahra Hamzavi, Birgitta Noone, and Kai Rosman from Stockholm University for their support. Experimental field work received financial support from the ERC under grant ERC-CoG-615922-BLACARAT and from the access support scheme of the Research Council Norway, project number 291644, "Svalbard Integrated Arctic Earth Observing System – Knowledge Centre, operational phase". The aerosol observations at Zeppelin observatory were supported also by Swedish Environmental Protection Agency (Naturvårdsverket), Knut and Alice Wallenberg foundation (KAW) project ACAS (grant no. 2016.0024), Swedish Research Council (grant no. 2018-05045) and ACTRIS-Sweden. B.B. received funding from the Swiss National Science Foundation (grant no 200021_204823) and from the European Union's Horizon 2020 research and innovation programme under the Marie Skłodowska-Curie grant agreement No 884104 (PSI-FELLOW-III-3i).

**Author contributions.**
BB: methodology, formal analysis, software, data curation, visualization, writing – original draft.
RLM: Funding acquisition, conceptualization, methodology, supervision, investigation, formal analysis, writing – review & editing.
GPF: investigation, writing – review & editing
RK: Funding acquisition, conceptualization, investigation, methodology, writing – review & editing
REP: investigation, writing – review & editing
PZ: conceptualization, investigation, methodology, writing – review & editing

MGB: Funding acquisition, conceptualization, methodology, supervision, project administration, writing – review & editing.

**Competing interests.**

The authors declare no competing interests.

**Code availability.**

Upon acceptance for publication the following repository will be made public on Zenodo (each item will have a separate DOI):
1. Currently, the SP2-XR analysis code can be found at [https://gitea.psi.ch/APOG_public/SP2XR](https://gitea.psi.ch/APOG_public/SP2XR). Upon acceptance of the manuscript, a DOI will be assigned to the version used to process the dataset used for this publication.
2. Python scripts for the data analysis and visualization related to this manuscript will be uploaded on Zenodo.

**Data availability.**

Upon acceptance for publication the following datasets will be made public on the Bolin Data Center:
a) SP2-XR parquet files. Converted, sorted, standardized files obtained from the SP2-XR analysis code (1).
b) Intermediate and final results relative to this manuscript including Figure raw data obtained from the analysis routines (2).

Temperature and wind speed data have been downloaded from Lund Myhre et al. (2025).

# Supplementary Information for
"Black Carbon scavenging in liquid Arctic clouds: the role of size and mixing state"

## 1 Properties of the sampled and selected clouds

Out of the 120 sampled cloud events, 37 meet all the filtering criteria. Figure S1 shows the distribution of cloud-averaged meteorological conditions and aerosol properties for all sampled cloud events (grey) and for the selected subset, categorized by BC scavenging type.

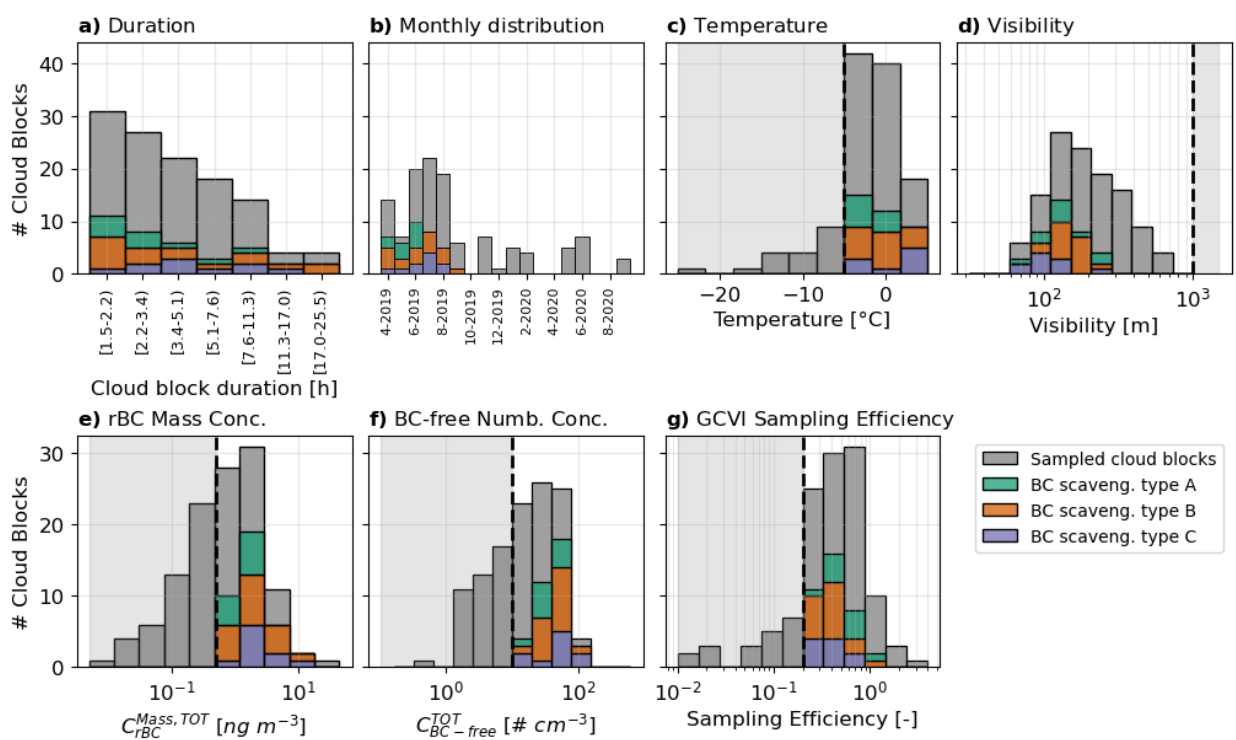

*Figure S1 Distribution of cloud-averaged properties for all sampled cloud events (grey) and for the selected events (colored bars, stacked by BC scavenging type). Vertical black dashed lines indicate the threshold values applied in the cloud event filtering criteria. The gray shaded areas highlight the range of values corresponding to cloud events excluded by the specific criterion.*

## 2 Cloud-averaged values for Figure 1

Figure 1 in the manuscript presents the variability in rBC mass size distribution and size-resolved BC scavenging, across and within different months. To complement these results, Figure S2 and Figure S3 show the normalized cloud-averaged rBC mass and number size distributions sampled from the total aerosol population for the selected clouds. Each distribution is normalized by its integral calculated over the rBC diameter range $D_{rBC} \in [68, 332)$ nm, allowing for a clearer comparison of distribution shapes across events. The curves are color-coded according to the corresponding BC scavenging group.

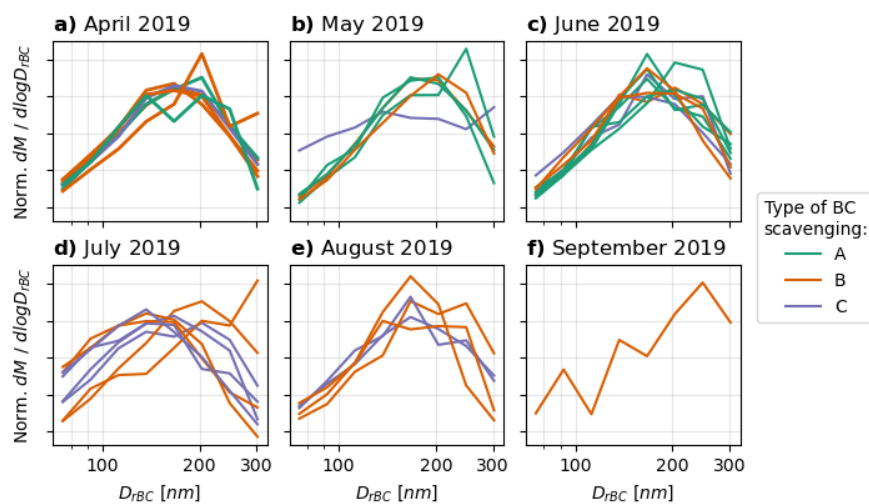

*Figure S2 Cloud-averaged rBC mass size distributions from the total aerosol population, normalized by the corresponding rBC mass concentration calculated over the rBC diameter range $D_{rBC} \in [68, 332)$ nm. Line colours indicate the BC scavenging type for each event.*

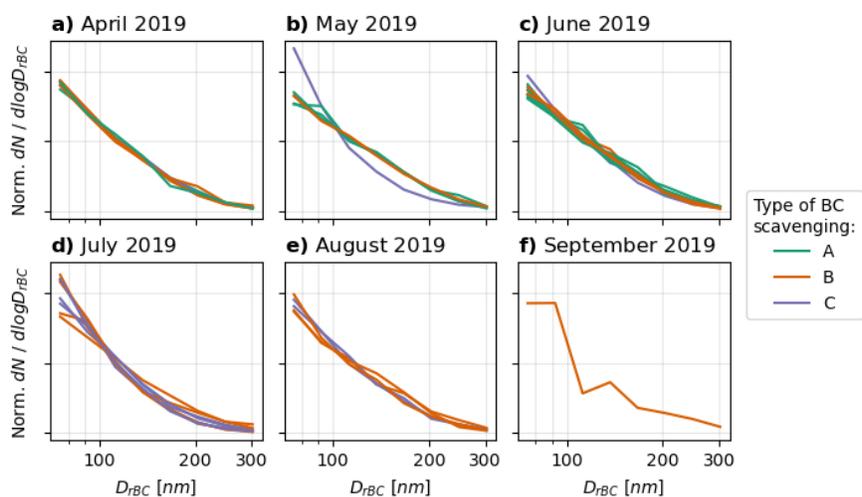

*Figure S3 Cloud-averaged BC particle number size distributions from the total aerosol population, normalized by the corresponding BC particle number concentration calculated over the rBC diameter range $D_{rBC} \in [68, 332)$ nm. Line colours indicate the BC scavenging type for each event.*

## 3  Classification of the size-resolved BC scavenging curves

The classification of BC scavenging types is based solely on the size-resolved BC scavenging curves. Cloud events with scavenged fractions above 0.9 in the diameter range $D_{rBC} \in [68,83)$ nm are classified as fully activated, indicating complete scavenging of BC particles without any size dependence (group A, 10 cloud events; see Figure S4a). The remaining 27 cloud events are further grouped using the K-means clustering algorithm from the scikit-learn Python library[1] applied to the size-resolved BC scavenging curves. This unsupervised classification yields two additional categories: group B (17 cloud events; Figure S4b), representing events with weak size dependence, and group C (10 cloud events; Figure S4c), characterized by strong size dependence.

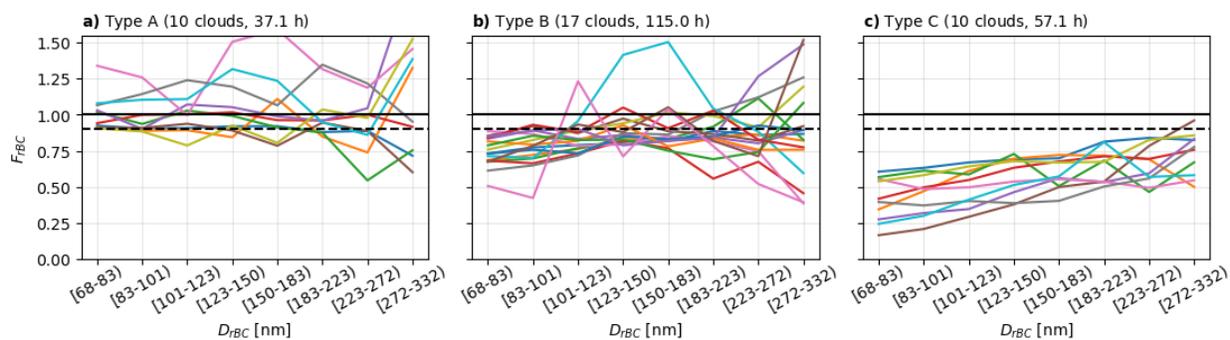

Figure S4 Cloud-averaged size-resolved BC scavenging curves for all selected cloud events, grouped according to the BC scavenging classification: (a) Group A: complete scavenging, (b) Group B: weak size dependence, and (c) Group C: strong size dependence.

## 4 Size-resolved scavenging of BC-free particles

Figure S5 shows the cloud-averaged scavenging curves for BC-free particles divided according to the BC scavenging classification. The areas highlighted with a grey shading indicate the diameter range used for the calculation of the GCVI sampling efficiency (see Section 6 below for more details). The average scavenged fraction in the [100, 150) nm optical diameter range corresponds to the data presented in the distribution plot in Figure 3d of the main manuscript. A clear distinction is observed across scavenging types: group A events show consistently high scavenging across all particle sizes, whereas groups B and C exhibit a lower average scavenged fractions in the optical diameter ranger 200 to 250 nm compared to type A events. However, even among group B and C events, individual cases with near-complete scavenging of BC-free particles are observed. Notably, the scavenged fractions of BC-free particles in this size range agree well with the BC scavenged fractions for thickly coated BC particles (black box plots in Figure 5) for all three cloud event types, thus indicating that tickly coated BC particles are as good CCN as BC-free particles with diameters well above the activation cut-off size.

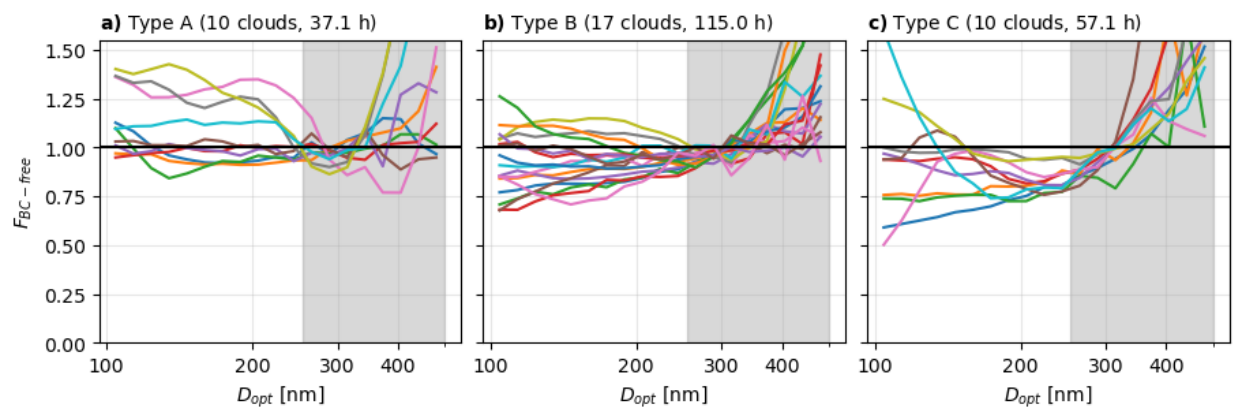

Figure S5 Cloud-averaged scavenging curves for BC-free particles, grouped according to the BC scavenging classification: (a) Group A: complete scavenging of BC particles, (b) Group B: weak size dependence in BC scavenging, and (c) Group C: strong size dependence in BC scavenging. Gray shaded areas represent the diameter range used for the calculation of the GCVI sampling efficiency (see Section 6 below).

# 5  Time delay analysis

The mixing state of BC particles can be qualitatively assessed using the time delay method[2]. This approach relies on the relative timing of signal peaks from elastic scattering and incandescence, which depends on BC mixing state. An internally mixed particle containing BC and non-refractory matter gradually heats up when entering the laser beam and evaporates in two distinct steps: first loosing non-refractory coatings, followed by BC sublimation with some time lag. Temporal evolution of the scattering and incandescent signals, in particular the time delay Δτ between their peaks, provides qualitative information on the amount of coating material and therefore the particle's mixing state. Due to the non-uniform intensity profile of the laser beam and the unknown position of the particle within it, this method does not allow for quantification of coating thickness (as opposed to techniques like the leading-edge-only fit method[7]).

The different mixing state classes used in this manuscript follow convention from previous literature studies[3–6]. When the peaks of incandescence and scattering signals occur almost simultaneously (small time delay) we classify the BC particle as uncoated to moderately coated, while larger time delays imply thicker coatings and thus a larger fraction of coating material (thickly coated). The threshold separating the coating categories corresponds to a coating volume fraction of approximately 70%[3]. Because the timing of the incandescence signal also depends on particle mass, the classification is applied in a size-resolved manner. BC particles with saturated scattering signal are assigned to the thickly coated category because the scattering signal from BC at incandescence is well below the saturation limit.

In some cases, the scattering peak can occur after the incandescence peak, which results in negative time delay values. This occurs if a chunk of BC-free material separates from the BC as the particle heats up. This chunk does not further evaporate and can cause a peak scattering signal when reaching the centre of the laser beam, typically well after the incandescence peak. Such particle splitting occurs either for BC attached to rather than embedded in the non-refractory matter, or for extremely thick coatings[6]. We observed such negative time delays, which were associated with extremely thick coatings, and therefore assigned particles with negative time lags to the thickly coated BC category.

It is also possible that the scattering signal of BC particles does not exceed the lower limit of detection. At the maximal BC core size at which this occurs, i.e. at $D_{rBC}$ ≈ 100 nm for our instrument, vanishing scattering signal implies bare or almost bare BC (if moderate coating was present, the scattering signal would exceed the lower limit of detection). Therefore, we treat them as a subset of the uncoated to moderately coated category, referred to as bare or almost bare particles. For even smaller BC cores, the scattering signal may also vanish for thicker coatings. Therefore, we limit the mixing state analysis to the range $D_{rBC}$ > 100 nm.

The classification was validated by visual inspection of numerous raw signal traces. A summary of the criteria used to assign the particles to either of the two categories, or to discard them from analysis, is provided in Figure S6 for the example of one cloud event.

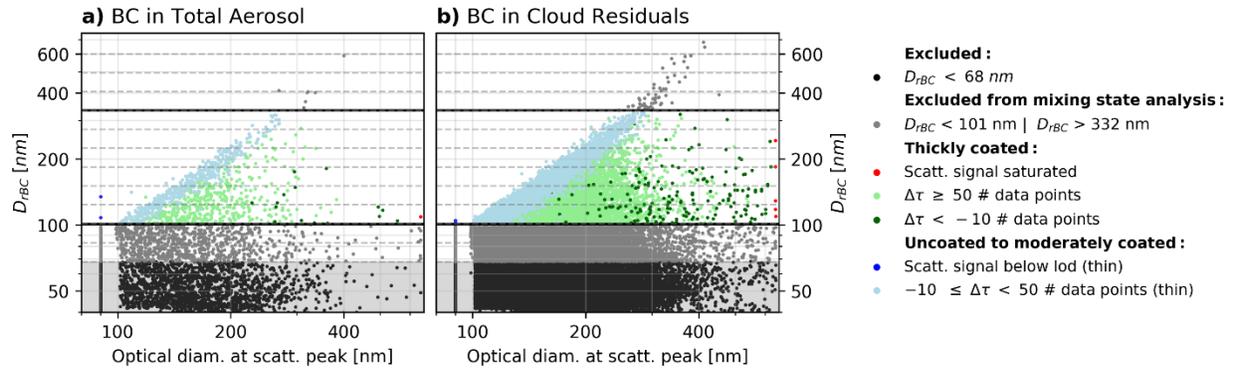

*Figure S6 Example of time delay classification at the single particle level. Particles with BC core diameter below 68 nm were excluded from all the analysis due to low counting efficiency, and those with diameter above 332 nm were excluded due to limited statistics. For the mixing state analysis, only particles with core diameters above 100 nm were considered, as smaller particles often lack a detectable scattering signal.*

# 6 GCVI Sampling Efficiency

The factors influencing the data collected with the Ground-based Counterflow Virtual Impactor (GCVI) inlet are the cut size (i.e., the size at which 50% of the droplets are sampled), the enrichment factor, and the size-dependent losses of cloud droplets in the main body of the GCVI. A detailed description of the GCVI working principle together with laboratory validation experiments is presented in Shingler et al.[8], while a detailed description of the GCVI and inlet setup used at the Zeppelin Observatory is presented in Karlsson et al.[9].

The cut size is defined by the operational conditions of the GCVI (i.e., air velocity in the wind tunnel and sample flow) and was calculated at ~6-7 µm for the measurements presented in this manuscript. The enhancement factor (EF) results from the concentration of the ambient air into the GCVI inlet and it is defined by the geometry of the GCVI and the air and sample flows. During the selected cloud events the cloud-averaged EF varied in the range 10.1 – 10.3. The sampling efficiency (SE) of the GCVI is mostly affected by the position of the droplet size distribution relative to the GCVI cut-off diameter, in addition to inertial deposition losses. To perform a theoretical estimate of the sampling efficiency, the cloud droplet size distribution is therefore needed. However, an indirect evaluation of the GCVI SE is also possible by comparing the total and cloud residual aerosol number size distributions[9]. In fact, by assuming that all the accumulation-mode aerosol particles are scavenged into cloud droplets, it is possible to estimate SE of the GCVI inlet from their ratio for liquid clouds. If this approach was applied to mixed-phase clouds, then the potential effect of the Wegener-Bergeron-Findeisen process on scavenged fractions would be eliminated from the measured scavenging results. Therefore, we focus on liquid clouds in this study.

The sampling efficiency for the cloud events we analyse in the manuscript is calculated by using the number size distribution of BC-free particles (i.e., particles that do not produce a detectable incandescence signal) measured by the SP2-XR from the total and GCVI inlets. We use BC-free particles in the diameter range from ~150 nm to ~500 nm and divide them in two size bins: $bin_1 = [153, 254)\ nm$ and $bin_2 = [254, 499)\ nm$. For each bin, the sampling efficiency is calculated as the ratio of the enhancement factor corrected GCVI concentration to the total aerosol concentration:

$$SE_{bin\ X} = \frac{C^{GCVI}_{BC-free, bin\ X}}{EF} \cdot \frac{1}{C^{TOT}_{BC-free, bin\ X}}$$

$SE_{bin\ X}$ represents the calculated sampling efficiency for $bin_x$, with $x = 1\ or\ 2$, $C^{GCVI}_{BC-free, bin\ X}$ is the concentration of BC-free particles sampled from the GCVI inlet, EF is the enhancement factor, and $C^{TOT}_{BC-free, bin\ X}$ is the concentration of BC-free particles sampled from the total inlet.

Since the SP2-XR was switching between the GCVI and total inlets every 10 minutes (switching cycle), we don't have collocated measurements of cloud residuals and total aerosol concentrations. We therefore calculate the SE to the shortest possible time scale of ~20 minutes. To improve the statistics of the aerosol concentration measured from the total inlet, we average two consecutive total inlet cycles and use them to calculate the SE with the average of the GCVI switching cycle in between them. Figure S7 graphically shows how the SE is calculated for a cloud event. Figure S7a shows the concentration of BC-free particles in $bin_2$ sampled by the GCVI (blue) and total (orange) inlets at 1-minute time resolution (small dots). For each inlet, we represent the averaging time with horizontal shaded areas and the resulting average value with larger dots. The resulting SE calculated for each 20-minute cycle is shown with black squares (right axis). Figure S7b shows the BC-free particle number size distribution obtained for the switching cycle indicated with a vertical dashed line in panel a. The blue size distributions represent the EF-corrected (dashed line) and EF-SE-corrected (solid line) cloud residual distributions. The calculated SE time series is then used to correct the corresponding 1-minute GCVI data. Figure S8 shows an example of how the SE correction is applied to the BC-free particle number concentration (panel a) and to the rBC mass concentration (panel b). The horizontal dashed lines represent the cloud-averaged concentrations that are then used to calculate the cloud-averaged bulk scavenged fractions.

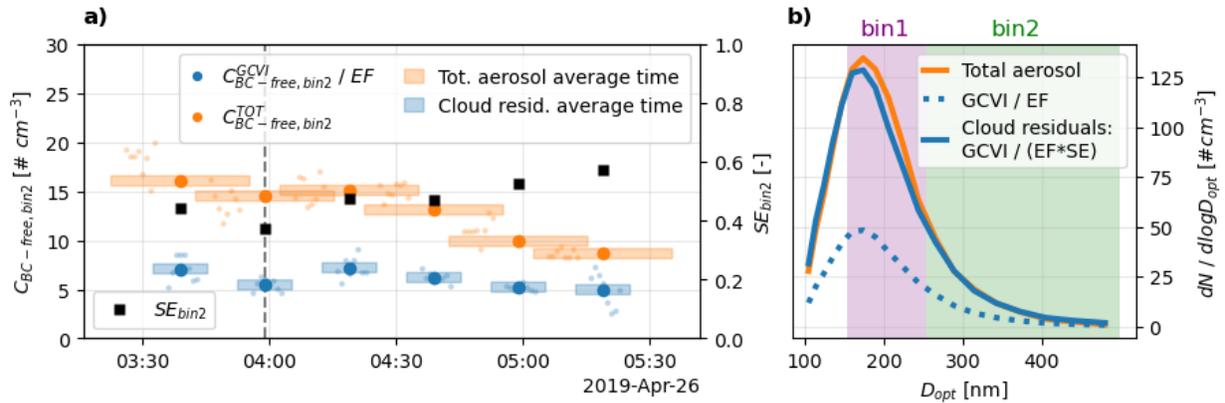

*Figure S7 Graphical representation of the sampling efficiency calculation when using the number concentration of BC-free particles in the size range of $bin_2$. Panel a: Number concentration time series of BC-free particles in the diameter range [254, 499) nm sampled from the total aerosol inlet (orange) and from the GCVI inlet (blue). Small dots represent 1-minute data, larger dots correspond to the switching cycle average (two 10-minutes periods for the total aerosol and one 10-minute period for the GCVI data). The horizontal bars represent the averaging period used for each corresponding large circle. The right axis shows the sampling efficiency correction factor obtained from the left axis data. Panel b: Number size distributions of BC-free particles from the total aerosol inlet (orange) and from the GCVI inlet (blue) with (solid line) and without (dashed line) sampling efficiency correction factor applied. The size distribution shown in panel b corresponds to the time stamp indicated with a vertical dashed line in panel a.*

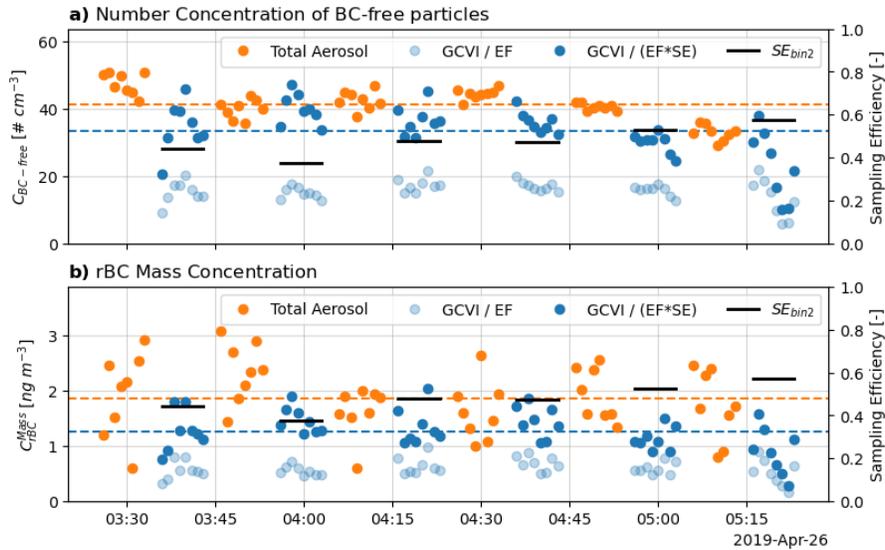

*Figure S8 Example time series of BC-free particle number concentration (panel a) and rBC mass concentration (panel b) sampled from the total (orange) and GCVI (blue) inlets. GCVI data are shown before (transparent) and after (solid) applying the sampling efficiency (SE) correction. The cloud average values are represented with horizontal dashed lines. The sampling efficiency values used to correct the 1-minute GCVI data are shown on the right axes with black bars.*

To validate the sampling efficiency values derived from the SP2-XR BC-free data, we compare our SE with values from previous studies that applied the same method. Karlsson et al.[9] used two differential mobility particle sizers (DMPS), while Pereira Freitas et al.[10] employed a multiparameter bioaerosol spectrometer and an optical particle counter. Figure S9a shows the comparison between the BC-free number concentration in the size range of $bin_2$ measured from the total aerosol inlet and from the GCVI inlet with the enhancement factor (EF) correction applied for the subset of selected cloud events. Each point in Figure S9a corresponds to a 20-minute time resolution data (i.e., a switching cycle). The best fit has a slope of 0.45, in agreement with the slopes obtained in the other two mentioned studies (0.45 and 0.46). Figure S9b shows the distribution of the calculated SE values for all the sampled clouds (grey) and for the subset of selected clouds (pink).

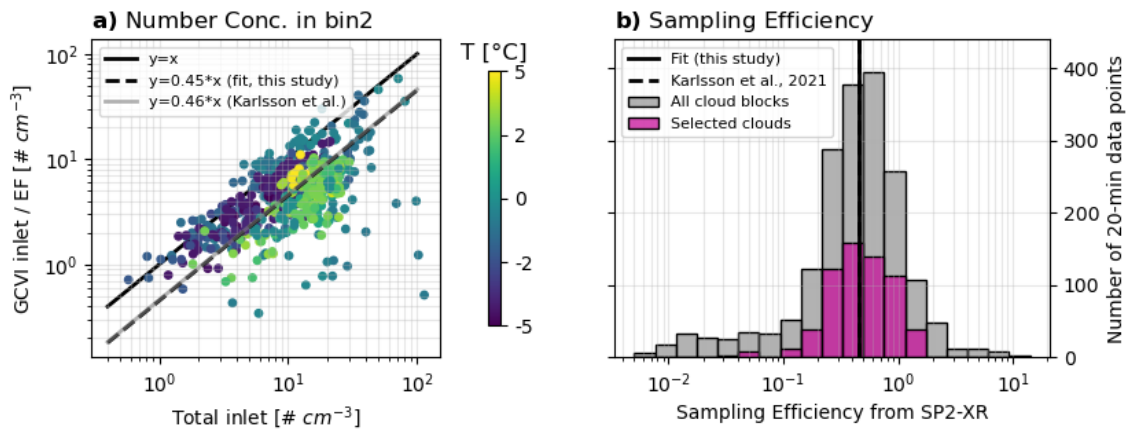

*Figure S9 Panel a: Comparison of BC-free particle number concentration in the size range of $bin_2 = [254, 499)$ nm measured from the total and GCVI inlets coloured by ambient temperature. Panel b: Distribution of the calculated sampling efficiency*

*values for all (grey) and selected (pink) cloud events. In both panels, the SE value obtained in a previous study performed at the Zeppelin Observatory with the same setup is shown for reference[9].*

# 7 Cloud selection

Among the sampled cloud events, defined as periods with visibility below 1 km and both the GCVI and SP2-XR in operation, we analyse only a subset to simplify data interpretation and ensure high data quality. The filtering criteria applied to identify the selected events are described and justified below.

1. Cloud-averaged temperature greater than -5 °C
   Tobo et al.[11] have shown that in the Arctic, some particles can initiate ice nucleation at temperatures as high as -5 °C. Clouds forming at temperatures below this threshold are therefore likely to be mixed-phase. By applying a temperature filter, we aim to select liquid cloud events. The presence of ice crystals can influence the GCVI sampling efficiency and complicate the interpretation of cloud microphysical processes[9,12]. In particular, the so-called Wegener-Bergeron-Findeisen process[13–15], where ice crystals growth at the expense of cloud droplets can lead to the evaporation of cloud droplets with the subsequent release of previously scavenged particles back into the interstitial aerosol population, making the interpretation of scavenged fractions more challenging.

2. Median rBC mass concentration in the range [68, 332) nm greater than 0.5 ng m$^{-3}$ in the total aerosol population
   Very low rBC concentrations can lead to noise amplification in the calculation of scavenged fractions. To avoid such uncertainties, we retain only cloud events with a median rBC mass concentration above 0.5 ng m$^{-3}$ in the total aerosol population, calculated from the 1-minute data for BC cores in the rBC diameter range $D_{rBC} \in [68, 332)$ nm. Figure S10a shows the distribution of these cloud-median concentrations for all sampled clouds (grey) and for those that satisfy this criterion (pink).

3. Median BC-free number concentration greater than 10 cm$^{-3}$ in the total aerosol population
   Extremely clean conditions can lead to inaccurate estimation of sampling efficiency and increase statistical noise. To reduce this effect, we retain only cloud events with a median BC-free number concentration above 10 cm$^{-3}$, calculated in the diameter range [100, 499) nm. Figure S10b presents the distribution of cloud-median BC-free number concentrations for all sampled clouds (grey) and for those satisfying this filter (pink).

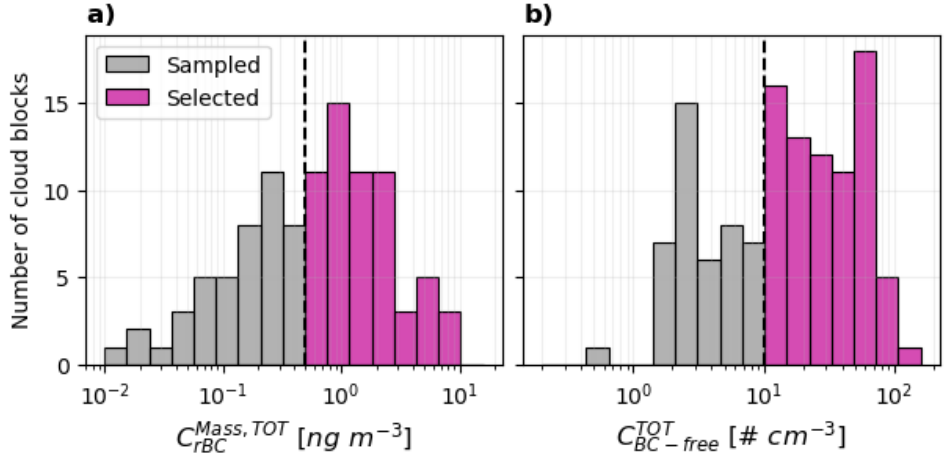

*Figure S10 Distribution of cloud-block median values of (a) rBC mass concentration for BC cores in the rBC diameter range $D_{rBC} \in [68, 332)$ nm and (b) BC-free number concentration in the diameter range [100, 499) nm. Grey bars represent all sampled cloud events, while pink bars indicate the events that meet the respective filtering criterion. The vertical dashed lines indicate the respective threshold values.*

4. Cloud-averaged GCVI sampling efficiency greater than 20%.
   This criterion is applied to exclude cloud events with low GCVI sampling efficiency, which could introduce high statistical uncertainty due to the poor representation of cloud droplets. Ensuring sufficient sampling efficiency improves the robustness of the derived scavenged fractions.
5. Relative difference between sampling efficiencies (SE) in the diameter ranges $bin_1 = [153, 254)$ $nm$ and $bin_2 = [254, 499)$ $nm$ is less than 0.5

$$\frac{|SE_{bin1} - SE_{bin2}|}{min(SE_{bin1},\ SE_{bin2})} < 0.5$$

Substantially different cloud-averaged SE values between $bin_1$ and $bin_2$ may indicate incomplete scavenging of accumulation-mode particles, which violates the assumption used in the SE calculation. This criterion may also exclude cloud events characterized by a low peak supersaturation with a consequent high critical diameter and a more size-dependent BC scavenging, or cloud events with large changes in cloud properties on timescales shorter than the switching time (~10 minutes) that can lead to higher measurement uncertainty. However, quantitative interpretation of scavenging for these clouds would require an independent quantification of SE. Figure S11a shows the distribution of the relative differences between SE values calculated in $bin_1$ and $bin_2$ for all sampled cloud events (grey) and for the subset of clouds that satisfy this specific criterion (pink). The vertical dashed line in Figure S11a marks the threshold of 0.5. Figure S11b provides two examples of cloud-averaged BC-free particle number size distributions for the total inlet and for the GCVI data corrected for enrichment factor (EF) and sampling efficiency (SE).

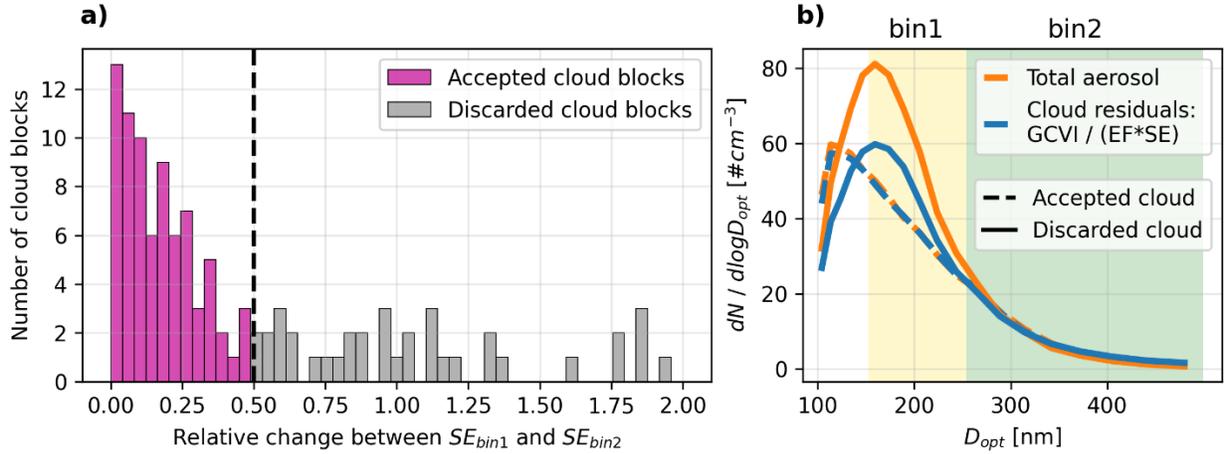

*Figure S11 **Panel a**: Distribution of the relative difference between the cloud-averaged sampling efficiency (SE) values calculated in $bin_1 = [153, 254)\ nm$ and $bin_2 = [254, 499)\ nm$. The vertical dashed line indicates the 0.5 threshold used to ensure that most accumulation-mode particles were scavenged into cloud droplets. Pink bars represent the cloud events that satisfy this specific filtering criteria. **Panel b**: Two examples of cloud-averaged BC-free number size distributions for individual cloud events. The event shown with solid line was discarded due to a large discrepancy between the total (orange) and GCVI-corrected (blue) data. The event shown with dashed lines was retained.*

## 8 Cloud-averaged values for the selected cloud events

The table below summarizes some of the mean or median parameters relative to the selected cloud events analysed in the manuscript. Temperature, visibility and GCVI sampling efficiency (SE) are cloud-averaged values. Columns that refer to BC sampled from the total aerosol population are shaded in grey.

| ID | Cloud Group | Start | End | Duration [h] | T [°C] | Visib. [m] | SE [-] | rBC Mass Conc | | | | Lognorm fit $dM/dlogD_{rBC}$ | | | |
|---|---|---|---|---|---|---|---|---|---|---|---|---|---|---|---|
| | | | | | | | | Mean | | Median | | Modal D [nm] | | GSD [-] | |
| | | | | | | | | Total | Cloud Res. | Total | Cloud Res. | Total | Cloud Res. | Total | Cloud Res. |
| 3 | B | 2019-04-16 07:51:00 | 2019-04-17 04:10:00 | 20.3 | -1.8 | 145 | 0.75 | 3.5 | 2.9 | 2.1 | 1.6 | 157 | 162 | 1.7 | 1.6 |
| 4 | B | 2019-04-17 17:03:00 | 2019-04-18 08:38:00 | 15.6 | -1.1 | 88 | 1.01 | 4.8 | 3.9 | 3.8 | 2.9 | 168 | 172 | 1.7 | 1.7 |
| 5 | A | 2019-04-18 10:20:00 | 2019-04-18 12:11:00 | 1.9 | -0.6 | 70 | 1.26 | 2.7 | 2.4 | 2.4 | 2.2 | 168 | 166 | 1.7 | 1.7 |
| 6 | C | 2019-04-18 13:37:00 | 2019-04-18 19:18:00 | 5.7 | -0.4 | 237 | 0.41 | 11.6 | 8.4 | 8.2 | 6.1 | 166 | 180 | 1.7 | 1.7 |
| 7 | B | 2019-04-18 20:03:00 | 2019-04-19 06:48:00 | 10.8 | -0.6 | 98 | 0.40 | 9.8 | 8.3 | 8.1 | 5.6 | 159 | 178 | 1.7 | 1.7 |
| 14 | B | 2019-04-26 03:26:00 | 2019-04-26 05:23:00 | 2.0 | -4.0 | 168 | 0.44 | 1.9 | 1.3 | 1.9 | 1.3 | 196 | 174 | 1.9 | 1.7 |
| 15 | A | 2019-04-26 13:31:00 | 2019-04-26 16:19:00 | 2.8 | -4.0 | 119 | 0.63 | 1.0 | 0.8 | 0.8 | 0.9 | 176 | 164 | 1.9 | 1.7 |
| 16 | C | 2019-05-13 13:40:00 | 2019-05-13 18:40:00 | 5.0 | -2.3 | 151 | 0.74 | 1.3 | 0.8 | 1.1 | 0.6 | 157 | 167 | 2.4 | 1.7 |
| 18 | A | 2019-05-15 05:11:00 | 2019-05-15 07:20:00 | 2.2 | -3.9 | 214 | 0.44 | 2.0 | 1.7 | 1.9 | 1.6 | 204 | 178 | 1.8 | 1.7 |
| 19 | B | 2019-05-15 07:49:00 | 2019-05-16 05:55:00 | 22.1 | -4.7 | 182 | 0.49 | 0.9 | 0.8 | 0.7 | 0.6 | 179 | 179 | 1.7 | 1.7 |

| | | | | | | | | | | | | | | |
|---|---|---|---|---|---|---|---|---|---|---|---|---|---|---|
| 20 | A | 2019-05-16 07:45:00 | 2019-05-16 17:02:00 | 9.3 | -4.6 | 98 | 0.79 | 0.9 | 0.9 | 0.8 | 0.9 | 178 | 176 | 1.7 | 1.7 |
| 21 | A | 2019-05-16 19:02:00 | 2019-05-16 21:10:00 | 2.1 | -4.6 | 86 | 0.74 | 1.1 | 1.1 | 1.1 | 1.1 | 171 | 182 | 1.6 | 1.7 |
| 22 | B | 2019-05-17 02:42:00 | 2019-05-17 04:55:00 | 2.2 | -4.6 | 202 | 0.67 | 1.9 | 1.7 | 1.9 | 1.8 | 222 | 189 | 1.9 | 1.7 |
| 23 | C | 2019-06-03 11:57:00 | 2019-06-03 15:01:00 | 3.1 | -2.9 | 94 | 0.50 | 1.2 | 0.7 | 1.0 | 0.5 | 186 | 169 | 1.8 | 1.7 |
| 24 | B | 2019-06-03 16:05:00 | 2019-06-03 17:36:00 | 1.5 | -3.8 | 131 | 0.44 | 2.7 | 2.1 | 2.3 | 2.1 | 189 | 169 | 1.8 | 1.6 |
| 25 | A | 2019-06-03 21:28:00 | 2019-06-04 00:14:00 | 2.8 | -4.5 | 168 | 0.62 | 1.4 | 1.2 | 1.1 | 0.8 | 170 | 170 | 1.6 | 1.6 |
| 33 | C | 2019-06-12 13:16:00 | 2019-06-12 23:31:00 | 10.3 | -2.4 | 64 | 0.49 | 6.1 | 3.8 | 5.6 | 3.3 | 158 | 173 | 1.8 | 1.6 |
| 34 | A | 2019-06-13 01:20:00 | 2019-06-13 02:55:00 | 1.6 | -2.1 | 211 | 0.35 | 1.0 | 1.2 | 0.8 | 1.2 | 194 | 175 | 1.8 | 1.6 |
| 35 | A | 2019-06-13 05:51:00 | 2019-06-13 13:04:00 | 7.2 | -1.1 | 142 | 0.34 | 1.2 | 1.4 | 1.2 | 1.4 | 173 | 178 | 1.7 | 1.7 |
| 36 | A | 2019-06-13 22:52:00 | 2019-06-14 03:03:00 | 4.2 | -0.9 | 137 | 0.34 | 1.7 | 1.6 | 1.6 | 1.6 | 174 | 196 | 1.7 | 1.9 |
| 38 | A | 2019-06-22 19:58:00 | 2019-06-22 23:06:00 | 3.1 | 0.9 | 114 | 0.27 | 1.7 | 1.9 | 1.2 | 1.1 | 193 | 188 | 1.7 | 1.8 |
| 39 | B | 2019-06-23 01:07:00 | 2019-06-23 04:05:00 | 3.0 | 0.6 | 126 | 0.27 | 5.0 | 4.5 | 5.1 | 4.3 | 158 | 178 | 1.6 | 1.6 |
| 40 | B | 2019-06-23 04:41:00 | 2019-06-23 07:43:00 | 3.0 | 1.0 | 161 | 0.24 | 3.1 | 2.9 | 3.1 | 2.8 | 172 | 179 | 1.7 | 1.7 |
| 45 | C | 2019-07-10 11:47:00 | 2019-07-10 23:38:00 | 11.9 | 3.9 | 104 | 0.39 | 1.0 | 0.5 | 0.7 | 0.4 | 171 | 209 | 1.9 | 1.8 |
| 46 | B | 2019-07-10 23:59:00 | 2019-07-11 02:11:00 | 2.2 | 2.8 | 196 | 0.28 | 0.7 | 0.7 | 0.5 | 0.5 | 300 | 182 | 2.3 | 1.6 |
| 54 | B | 2019-07-18 02:59:00 | 2019-07-18 13:20:00 | 10.4 | 0.6 | 148 | 0.33 | 1.6 | 1.3 | 1.3 | 1.2 | 188 | 202 | 1.7 | 1.7 |
| 62 | C | 2019-07-23 12:24:00 | 2019-07-23 15:47:00 | 3.4 | 2.9 | 75 | 0.33 | 1.3 | 0.5 | 1.1 | 0.3 | 135 | 212 | 1.9 | 1.8 |
| 63 | C | 2019-07-30 06:17:00 | 2019-07-30 08:19:00 | 2.0 | 5.6 | 109 | 0.63 | 1.8 | 1.0 | 1.6 | 0.9 | 152 | 156 | 1.7 | 1.7 |
| 64 | B | 2019-07-30 16:48:00 | 2019-07-30 19:45:00 | 3.0 | 4.8 | 166 | 0.31 | 1.6 | 1.3 | 1.4 | 1.3 | 128 | 130 | 1.7 | 1.7 |
| 65 | B | 2019-07-30 20:57:00 | 2019-07-31 04:20:00 | 7.4 | 3.6 | 116 | 0.39 | 1.1 | 0.9 | 1.1 | 0.9 | 129 | 126 | 1.6 | 1.7 |
| 66 | C | 2019-07-31 05:14:00 | 2019-07-31 09:55:00 | 4.7 | 3.0 | 113 | 0.23 | 3.2 | 1.4 | 2.3 | 1.3 | 130 | 145 | 1.7 | 1.9 |
| 67 | C | 2019-08-01 09:30:00 | 2019-08-01 18:20:00 | 8.8 | 3.4 | 104 | 0.28 | 1.3 | 0.9 | 1.2 | 0.8 | 188 | 191 | 2.0 | 1.8 |
| 68 | B | 2019-08-03 18:13:00 | 2019-08-03 22:31:00 | 4.3 | 2.6 | 156 | 0.35 | 1.5 | 1.4 | 1.4 | 1.4 | 197 | 188 | 1.7 | 1.7 |
| 73 | C | 2019-08-09 07:22:00 | 2019-08-09 09:40:00 | 2.3 | 2.4 | 117 | 0.30 | 2.4 | 1.3 | 2.2 | 1.0 | 165 | 181 | 1.8 | 1.5 |
| 74 | B | 2019-08-11 18:44:00 | 2019-08-11 20:45:00 | 2.0 | 0.0 | 134 | 0.28 | 1.7 | 1.5 | 1.6 | 1.4 | 161 | 169 | 1.5 | 1.6 |
| 75 | B | 2019-08-13 09:54:00 | 2019-08-13 13:43:00 | 3.8 | -0.3 | 141 | 0.31 | 0.7 | 0.6 | 0.6 | 0.5 | 178 | 167 | 1.9 | 1.7 |
| 90 | B | 2019-09-22 11:26:00 | 2019-09-22 12:56:00 | 1.5 | -4.2 | 216 | 0.51 | 0.5 | 0.4 | 0.5 | 0.4 | 273 | 184 | 2.3 | 1.6 |